\documentclass[	DIV=calc,%
							paper=a4,%
							fontsize=11pt%
							]{scrartcl}	 					
\usepackage{lipsum}													

\usepackage[english]{babel}										
\usepackage[protrusion=true,expansion=true]{microtype}				
\usepackage{amsmath,amsfonts,amsthm}					
\usepackage[pdftex]{graphicx}									
\usepackage[svgnames]{xcolor}									
\usepackage[hang, small,labelfont=bf,up,textfont=it,up]{caption}	
\usepackage{epstopdf}												
\usepackage{subfig}													
\usepackage{booktabs}												
\usepackage{fix-cm}													

\usepackage{sectsty}													
\allsectionsfont{
	\usefont{OT1}{phv}{b}{n}
	}

\sectionfont{
	\usefont{OT1}{phv}{b}{n}
	}

\usepackage{fancyhdr}												
\usepackage{lastpage}	
\usepackage{lettrine}
\newcommand{\initial}[1]{%
     \lettrine[lines=3,lhang=0.3,nindent=0em]{
     				\color{DarkGoldenrod}
     				{\textsf{#1}}}{}}

\usepackage{titling}															

\newcommand{\HorRule}{\color{DarkGoldenrod}
									  	\rule{\linewidth}{1pt}%
										}

\pretitle{\vspace{-30pt} \begin{flushleft} \HorRule 
				\fontsize{20}{20} \usefont{OT1}{phv}{b}{n} \color{DarkRed} \selectfont 
				}
\title{2018 NASA Laboratory Astrophysics Workshop:\\
Scientific Organizing Committee Report}					
\posttitle{\par\end{flushleft}\vskip 0.5em}

\author{}											

\date{}																				

\begin{document}
\maketitle
\thispagestyle{fancy} 			

\noindent \today\\

\noindent Prepared by:

\vskip 2pc
\noindent N. Brickhouse (Smithsonian Astrophysical Observatory)\\
\\
G. J. Ferland (University of Kentucky)\\
\\
S. Milam (NASA Goddard Space Flight Center)\\
\\
E. Sciamma-O'Brien (NASA Ames Research Center)\\
\\
A. Smale (NASA Goddard Space Flight Center)\\
\\
A. Spyrou (Michigan State University)\\
\\
P. C. Stancil (University of Georgia)\\
\\
L. Storrie-Lombardi (Jet Propulsion Laboratory)\\
\\
G. M. Wahlgren (GDIT/STScI)\\
\\

\noindent External Reviewer: Randall Smith (Smithsonian Astrophysical Observatory)
\vskip 2pc
\noindent Special thanks to: David A. Neufeld and Susanna Widicus-Weaver

\newpage

\section{Executive Summary}

\initial{T}his
report provides detailed findings on the critical laboratory astrophysics data needs that are required to maximize the scientific return for NASA's current and near-term planned astrophysics missions. It also provides prioritized rankings on said laboratory astrophysics data, generally by waveband. The Report is based on community input gathered at the 2018 NASA Laboratory Astrophysics
Workshop (LAW) from presentations, from discussions during workshop breakout sessions, and from other solicited input deemed appropriate by the Scientific Organizing Committee (SOC) obtained prior to and after the meeting. Hence, the Report is a direct reflection of the spirit and participant make-up of LAW 2018. 
The Report also outlines specific opportunities and threats facing NASA's Laboratory Astrophysics Program, and articulates concrete actions by which the Agency can capitalize on the opportunities and mitigate the challenges.
Due to the upcoming preparations for the 2020 Decadal Survey (DS) for Astronomy and Astrophysics, an Appendix is included summarizing meeting discussions to solicit
laboratory astrophysics community input.
The Report was prepared by the SOC, with help from some invited speakers, and input and review from community members.

\subsection{Summary of Recommendations}

Highlights of key recommendations related to operating and approved future NASA missions include:
\begin{itemize}
\item Spectral studies for molecular line lists, especially focusing on excited vibrational states, isotopologues, and intensities for weaker lines, as well as extensions to $>$500 GHz;
\item Inner-shell cross section measurements of K- or L-shell ionization of atoms and ions, both by
  photons and electrons; 
\item Spectral and physical characterization of analogs of carbon and silicon grains (optical constants, morphology, composition of grains);
\item UV-IR spectra of PAHs, PANHs, large PAHs, large PANHs, PAH clusters, PAH derivatives, and fullerenes; and
\item Laboratory measurements and theory for nuclear reaction rates, half-lives, mass measurements, and atomic opacities for neutron-star mergers, kilonovae, and multi-messenger astronomy.
\end{itemize}
In addition, more general recommendations include:
\begin{itemize}
\item The creation of a US laboratory astrophysics institute or network,
\item The establishment of a funding mechanism to ensure continual support of laboratory astrophysics databases and related software tools/code packages,
\item Coordination with US and international experimental user facilities for continued access for laboratory astrophysics relevant experiments, and
\item The preparation of white papers for the 2020 Decadal Survey for Astronomy and Astrophysics by the laboratory astrophysics community.
\end{itemize}

\section{Charter and Scope of Meeting}

The purpose of the LAW 2018 was to identify and prioritize critical laboratory astrophysics data needs, defined
broadly, to meet the demands of NASA's current and near-term astrophysics missions. To that end,
the meeting attempted to provide a forum within which the community could present and review the current state of knowledge
in laboratory astrophysics and identify challenges and opportunities for the field in the near future.

The specific goals of LAW 2018 included:
\begin{enumerate}
\item Identify the critical laboratory astrophysics data needs of NASA's current and near-term planned astrophysics missions;
\item Prioritize those data needs based on the demand to maximize mission scientific return;
\item Assess the degree to which NASA-supported research efforts currently address the prioritized data needs;
\item Assess how funding of critical data needs may be shared across other agencies, e.g. NSF, DoE;
\item Review the current state-of-the art in laboratory astrophysics in general and as related to the prioritized data needs;
\item Review the recommendations of previous LAWs and assess progress toward meeting those recommendations, as appropriate;
\item Assess the strengths, weaknesses, opportunities, and threats facing NASA's Laboratory Astrophysics program in the context of the Astro2010 Decadal Survey report;
\item Begin developing strategies for community input to the Astro2020 Decadal Survey;
\item Generate an electronic volume of science proceedings from the workshop that will serve as a reference to NASA and the community (see www.physast.uga.edu/workshops/law/presentations); and
\item Formulate a report summarizing critical and prioritized laboratory astrophysics data needs which can be used as a guide for NASA  mission-driven funding of the field.
\end{enumerate}
The current document is the response to these goals, particularly item 10. However, it should not be misconstrued as a replacement for peer review and its recommendations may evolve due to changing mission needs.
The primary NASA program target for LAW 2018 and the Report is the Astrophysics Research and Analysis (APRA) program, Laboratory Astrophysics category. However, portions of the Report, specifically on Software Instruments, Databases, and Archives, are relevant to the Astrophysics Database research area for the Astrophysics Data Analysis Program (ADAP). 

\section{Current and Future Needs}

In this section we detail findings from the LAW 2018 meeting in relation to key laboratory astrophysics needs and their astrophysical motivations. Of critical importance to aid in the selection of potential funding areas by APRA and ADAP, the Report provides in this section 
prioritizations of laboratory astrophysics data deemed most relevant to current and approved future NASA astrophysics missions. The prioritizations were arrived at through community input prior to, during, and after the LAW 2018 meeting including post-meeting surveys. These findings are given broadly by waveband classifications, though overlaps are clearly evident.

\subsection{UV/Optical/Mid-IR missions}
Many NASA astrophysics missions are encompassed in the wide spectral range from ultraviolet (UV) to mid-infrared (MIR) wavelengths as listed in the below table. These missions enable scientific investigations of a broad range of astrophysical environments from comets to planetary and stellar systems to the interstellar medium and galaxies.\\
\noindent\includegraphics[width=\textwidth]{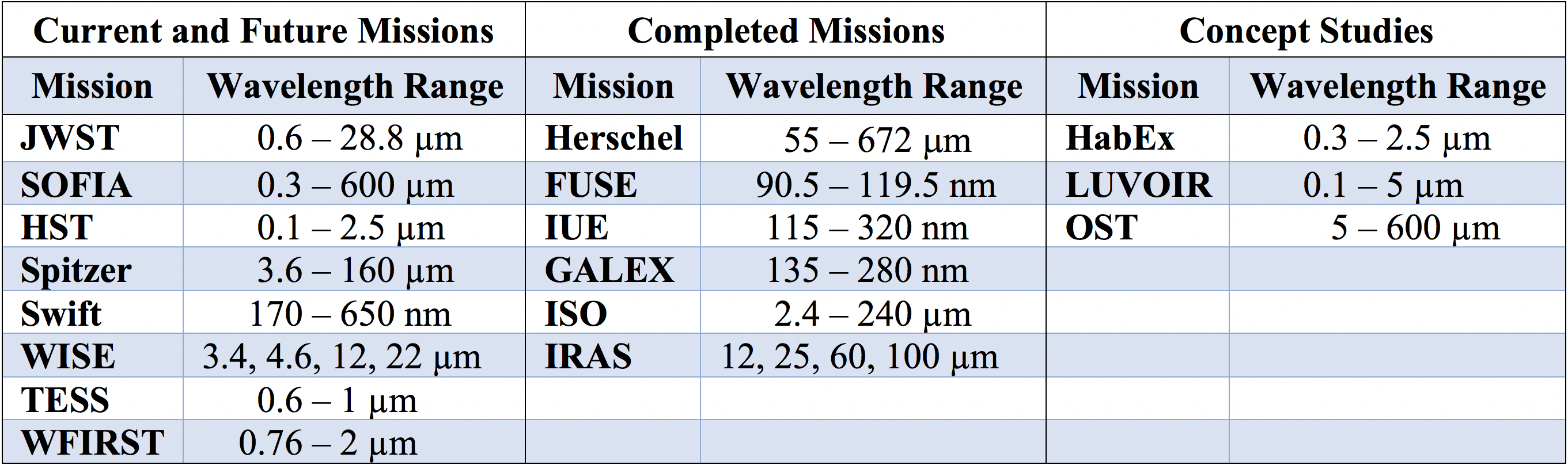}

\vspace{0.1cm}

During the UV/Optical/MIR break-out session of the LAW 2018, the discussion focused primarily on three areas, listed below, that were identified by the community as key areas in need of laboratory astrophysics data for the upcoming years:

\begin{enumerate}

\item The \textit{James Webb Space Telescope (JWST)} will measure the IR Universe across the entire NIR/MIR range (0.6-28.8 $\mu$m) with an unprecedented combination of high sensitivity, and spatial and spectral resolutions. \textit{JWST} will be a unique tool to study circumstellar matter, dust, the interstellar medium (ISM), molecular clouds and H II regions, in local and extragalactic sources. \textit{JWST} will be able to simultaneously observe and spatially resolve the spectral signatures of the ionized gas, atomic lines, all the known hydrocarbon and polycyclic aromatic hydrocarbon (PAH) bands, and dust and ice spectral features, hence giving access to diagnostics of the chemical and physical processes at play.

\item Exoplanetary science continues to rapidly evolve and laboratory astrophysics support for studies of exoplanet atmospheres, ices, and gases is critical.

\item While the next astrophysics flagship mission is an IR mission (\textit{JWST}), support for work in the UV is still important for the analysis of UV observational data from past and current missions, and studies of UV lines observed at high redshift in optical and near-IR (NIR) wavelengths, as well as the development of future NASA UV missions. 

\end{enumerate}

To exploit the data returned from NASA astrophysics missions to their full potential (analysis, interpretation, and understanding of mission data, as well as calibration of instruments, and prediction of detectable data to prepare for missions), many laboratory astrophysics measurements and calculations are needed. Because of the wide range of wavelengths encompassed in this section and the many applications and range of physical phenomena covered, the UV/Optical/MIR section has been divided into five subsections, following five core laboratory astrophysics disciplines: 
i) Atoms, ii) Molecules, iii) Dust, iv) Ice, and v) Exoplanets.
For each of these sections, the needs are listed in order of broad priority levels.

\vspace{0.2cm}

\subsubsection{Atoms}

NASA astrophysics missions are in critical need of reliable laboratory astrophysics data (theory and experiment) for key atomic properties  and atomic processes (dielectronic recombination (DR), electron impact ionization (EII), photoionization, etc.).

\vspace{0.3cm}

\noindent\includegraphics[width=\textwidth]{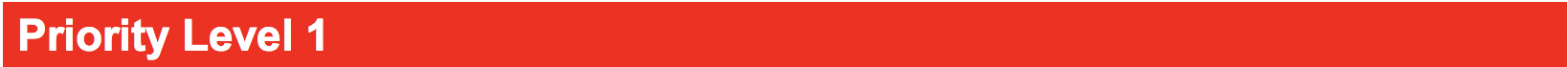}
\begin{itemize}

\item \textbf{Atomic data for atoms and ions (structure, wavelength, oscillator strengths, opacities, transition probabilities, energy levels) (experiment + theory)}\\
\textit{
- Currently available atomic data have large uncertainties for neutrals or first ions of a number of elements (especially Mg, Si, P, S, Cl, Ar, Ti, Co, Cu).\\
- Oscillator strengths are missing for many elements for wavelengths below about 180 nm and are necessary for the analysis of HST and FUSE data.  \\
- Improved oscillator strengths and opacities are critically needed for Cs, Te, Sb, I, Xe, Ce, Nd, Au, Pt, and many lanthanides, in order to model/interpret the observational spectra of neutron star mergers (NSMs).\\
- A range of other atomic line data are necessary to interpret UV/Optical observations of stars, the ISM, planetary nebulae (PNe), and gas in and around galaxies.
}

\item \normalsize{\textbf{Chemical kinetics (DR rates, photoabsorption, photoionization and EII cross sections, and collisional rate coefficients) (experiment + theory)}}\\
{\textit{- Understanding the spectral and thermal properties of cosmic atomic plasmas requires an accurate knowledge of the ionization level of the gas. Improved dielectronic recombination (DR) rates are essential for accurate determinations of ionization parameters and ionization corrections to elemental abundances. There are large uncertainties on currently available DR rates, particularly for elements beyond the second row of the periodic table.\\
- Calculations of collisional rate coefficients are needed to interpret SOFIA and future JWST observations of the fine-structure emission lines of atomic ions (e.g. [Fe II] 26 $\mu$m; [N II] 205 $\mu$m, etc. with SOFIA), which are an important diagnostic of, for example, active galactic nuclei (AGN) and star-forming environments.\\
- Gaps still exist in the measured and calculated photoionization cross sections, a parameter critically needed for interpreting data from NASA missions from the UV to the far IR (FIR).}}
\end{itemize}


\subsubsection{MOLECULES}
Molecules in astrophysical environments span from simple diatomics, to complex organic molecules (COMs), to PAHs. Because of the importance of PAHs, they are highlighted in a section of their own.\\

\centerline{\textbf{POLYCYCLIC AROMATIC HYDROCARBONS (PAHs)}}

\vspace{0.2cm}

\textit{JWST} will simultaneously provide long wavelength coverage (0.6-28.5 $\mu$m), medium spectral resolution, high spatial resolution, and high sensitivity. Such a unique combination is revolutionary as it will allow for a better decomposition of the PAH IR emission into its contributing components/features and will allow for the ability to probe the photochemical evolution of dust particles, very small grains, PAH clusters, fullerenes, and PAH molecules.

\vspace{0.3cm}
\noindent \includegraphics[width=\textwidth]{Priority_Level_1_ter.png}
\begin{itemize}
\item \textbf{UV-IR spectra of PAHs, PANHs, large PAHs, large PANHs, PAH clusters, PAH derivatives, and fullerenes (matrix and gas phase experiments + theory)}\\
\noindent \textit{- These laboratory astrophysics data are critical:\\
   $\rightarrow$ to understand the extinction curve: identification of the carriers of diffuse interstellar bands (DIBs), aromatic interstellar bands (AIBs),\\
   $\rightarrow$ to provide an accurate PAH emission model. Theoretical spectra need to be produced and then converted to emission spectra. Experimental spectra are needed to calibrate theoretical spectra of PAHs,\\
   $\rightarrow$ to understand the origin of the broad plateaus underlying the PAH features in the IR. \\
- Very large PAHs dominate the interstellar PAH family, but are poorly covered in the current PAH databases. This lack of large PAHs influences the interpretation of the astronomical IR observations which now relies on smaller PAHs, anionic PAHs, and/or protonated-PAHs.\\
- Several of the main interstellar PAH features cannot be matched exactly with pure PAH bands, but based on a few calculated species of nitrogen containing PAH cations, incorporation of N within the hexagonal C skeleton shifts these bands into agreement with the interstellar positions. No systematic study has been done on PANHs, while  investigations of large PANHs are completely missing.\\
- The overtone, combination, and hot-bands originating from highly-excited PAHs are not well probed, forcing reliance on the mostly 0~K absorption computed data currently available. There is also an ambiguity in contributions from PAH anions, supra-hydrogenated PAHs, and anharmonic effects to the extended red-wings on many of the astronomical PAH emission features. This ambiguity can only be resolved with proper anharmonic descriptions of the band profiles. Therefore, gas-phase experimental data and quantum chemical calculations including the effects of anharmonicity on or extrapolated to astronomically relevant PAH sizes are in dire need.\\
- Lab measurements of cold PAHs isolated in the gas phase and in matrices will enable computational chemists to ground truth their calculations for (i) very large PAHs and PANHs, and (ii) regarding anharmonicity, allowing them to adjust the theories as needed to produce high quality IR spectra of species that will likely never be studied in the lab.}

\item \normalsize{\textbf{Formation (reaction mechanisms to go from small molecules to PAHs), destruction and evolution of PAHs (photo- and shock-processing) (experiment + theory)}}\\
\textit{To understand how PAHs are formed and then processed in space by UV radiation, cosmic-ray bombardment, and shocks, calculations and measurements of chemical pathways, dissociation energies (and channels), relaxation rates, UV-MIR spectra and molecular structure of products are needed.}
\end{itemize}


\noindent \includegraphics[width=\textwidth]{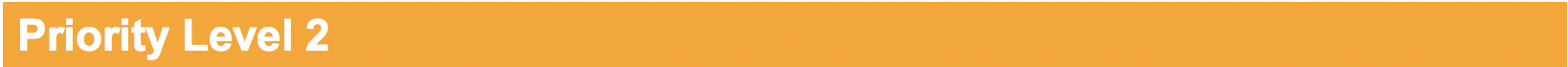}
\begin{itemize}
\item \textbf{H$_2$ formation on PAHs}\\
\textit{The H$_2$ formation efficiency as a function of PAH size and charge is an important parameter in many models. In order to quantify the role PAHs play in the formation of H$_2$ in the ISM, one first needs to know the size and charge distribution of the interstellar PAH family.}
\end{itemize}

\vspace{0.6cm}

\centerline{\textbf{MOLECULAR UNIVERSE}}

\vspace{0.3cm}

\noindent\includegraphics[width=\textwidth]{Priority_Level_1_ter.png}

\begin{itemize}
\item \textbf{Chemical kinetics and chemical sensitivity studies (rate coefficients, reaction products). Reaction network for small molecules leading to large molecular species and grains (experiment + theory)}\\
\textit{- To understand how large molecular species and grains are formed and then processed in space by UV radiation, cosmic ray bombardment, and shocks,  calculations and measurements of chemical pathways, rate coefficients, and molecular structure of products are needed.\\
- Isotopic anomalies and other signatures are carried by gas phase molecules and provide unique information on nucleosynthetic and astrochemical processes occurring in interstellar and protostellar environments.\\
- Branching ratios for photo-absorption processes.}

\item \normalsize \textbf{UV-MIR spectra of molecules in the gas phase (theory + experiment)}\\
\textit{Experimental and theoretical spectra of diatomic and polyatomic molecules are needed in the gas phase and the solid phase to identify spectral features in protostellar disks and star-forming regions.}
\end{itemize}


\noindent\includegraphics[width=\textwidth]{Priority_Level_2_ter.png}

\begin{itemize}
\item \textbf{Collisional rates with H$_2$, H, and He for molecules and ions observed in disks}\\
\footnotesize{\textit{- For all molecular transitions seen in disks at 1-28 $\mu$m (H$_2$O, OH, CH$_4$, CO$_2$, NH$_3$, CH$_3$OH, HCN, C$_2$H$_2$, ... and hydride isotopologues), and for other environments, molecular ions, such as HCO$^+$, CF$^+$.\\
- Needed for non-LTE modeling of emission features.}}

\end{itemize}


\subsubsection{DUST}

\noindent\includegraphics[width=\textwidth]{Priority_Level_1_ter.png}

\begin{itemize}
\item \textbf{Spectral and physical characterization of analogs of carbon and silicon grains (optical constants, morphology, composition of grains, crystalline \textit{vs.} amorphous, absorption features) (theory + experiment)}\\
\textit{- Essential to determine detailed grain properties (e.g., identify specific components and constrain grain sizes) from observations of MIR grain features from nearby and distant galaxies.\\
- Fundamental input for radiative transfer and microphysical models.\\
- Spectral features produced by mineral dust grains are needed to help identify the composition and abundance of materials in multiple astrophysical environments including protostellar disks and star formation regions.}

\item \normalsize{\textbf{Chemical reaction mechanisms and pathways studies, and heterogeneous catalysis, with and without photo-processing (theory + experiment)}}\\
\textit{- To study the evolution from small molecules to grains, then grain growth (sticking probabilities, microgravity, ...) and destruction (irradiation, shocks, sputtering, ...).\\
- To investigate surface-induced chemistry, mineral/organics/ice interactions.\\
- Dust and ice are largely responsible for H$_2$ and production of many complex organic molecules (COMs).}

\end{itemize}

\noindent\includegraphics[width=\textwidth]{Priority_Level_2_ter.png}

\begin{itemize}

\item \textbf{Dust polarization (experiment + theory)}\\
\textit{- To better understand grain alignment properties in dense molecular clouds. Polarization provides information on magnetic field alignments.  Experimental and theoretical study making connection with polarization strength and composition of grains is important.\\
- To give unique insight in dust grain composition. Most silicate absorption features appear amorphous. However, spectropolarimetry can show polarized components that allow the identification of specific silicates.}

\end{itemize}

\vspace{0.3cm}

\subsubsection{ICES}

\noindent\includegraphics[width=\textwidth]{Priority_Level_1_ter.png}

\begin{itemize}
\item \textbf{Ice kinetics, evolution/processing  (experiment + theory)}\\
\textit{- Deposition, desorption (photo + thermal), photodissociation, photoprocessing under UV irradiation/cosmic rays/x-rays, restructuring, chemistry, ice-grain interactions, chemical processing that causes isotopic fractionation.\\
- Dust and ice are largely responsible for H$_2$ and COM production.}

\item \normalsize{\textbf{Spectral characterization of interstellar ice analogs (IR spectra, optical constants, with and without irradiation, over a range of temperatures)}}\\
\textit{There are many gaps in the current databases, even for simple molecules and binary, tertiary mixtures (those commonly observed). Further characterization of ice properties, including spectral features for complex organic materials in icy matrices are needed.}

\end{itemize}

\noindent\includegraphics[width=\textwidth]{Priority_Level_2_ter.png}

\begin{itemize}
\item \textbf{Isotopic signatures of ices}\\
\textit{To interpret observational data of environments like dense molecular clouds, measurements of isotopic signatures from gas phase molecules are not enough as the main carriers of isotopic signatures are grains. The observation of gas phase molecules only does not give a good representation of the environment. For example, the ammonia D/H ratio is not representative of the overall cloud. The main reservoir in dense clouds is grains and ices and they might have a different D/H ratio.}

\end{itemize}

\vspace{0.3cm}

\subsubsection{EXOPLANETS}

\noindent\includegraphics[width=\textwidth]{Priority_Level_1_ter.png}

\begin{itemize}

\item \textbf{Spectral and physical characterization of organics (molecules + ices + aerosols) (experiment + theory)}\\
\textit{- To simulate gas and solid phase material relevant to exoplanetary atmospheres and surfaces and their interaction.}

\item \textbf{Chemical reaction rates relevant to photochemistry at high temperature} (different mixtures than typically studied for ISM or solar system applications)

\item \textbf{Line lists (molecular line lists at high temperature, rovibrational line lists for non-hydrid molecules up to 2000 K, line lists for biosignature and techno-signature molecules at low temperature) (theory + experiment)}\\
\textit{- To model opacity of molecular gases for comparison to exoplanet atmosphere observations (H$_2$, CO, H$_2$O, N$_2$, CH$_4$, NH$_3$, CO$_2$, HCN, H$_2$S, HS,... + non-hydrid species for hot, rocky exoplanets).\\
- For the study (observation and modeling) of the opacity of hot rocky exoplanets, and cool ($\sim$300 K) habitable rocky planets.}

\end{itemize}


\noindent\includegraphics[width=\textwidth]{Priority_Level_2_ter.png}

\begin{itemize}

\item \textbf{H$_2$ and He pressure broadening of molecular spectra at high temperatures} (for many of the molecular species listed above that are relevant to exoplanet atmospheres)

\item \textbf{Line lists of transition-metal diatomics}
\end{itemize}

\subsection{FIR/submm missions}
There are a limited number of FIR to sub-millimeter (submm) missions that are relevant for current and future needs of laboratory data; however, the data needs are quite extensive.  

\subsubsection{Post-Herschel Laboratory Astrophysics}
The \textit{Herschel Space Observatory} exhausted its helium and ended the mission in April 2013 and the \textit{Spitzer Space Telescope} ended its cryogenic mission in May 2009.  The last NASA LAW, LAW 2010, was held during the prime \textit{Herschel} mission and emphasis was placed on line lists of ``interstellar weeds" in order to search for new species.  The extensive data sets of these two missions are still being analyzed with publications forthcoming. Laboratory measurements of new molecules at these wavelengths are used in searching these datasets to identify the vast number of unknown spectral features present
.  Additionally, ESA recently selected the \textit{SPICA} M5 mission concept as one of the three studies for their Cosmic Visions program that will be finalized in 2021.  This is a cold FIR mission ranging from 12-230 $\mu$m, providing useful overlap with SOFIA, \textit{JWST}, \textit{Spitzer}, and \textit{Herschel}. Finally, one of the four astrophysics mission concept studies for consideration in the 2030s is the \textit{Origins Space Telescope (OST)} that will be a MIR to FIR mission that complements all of these observatories with unprecedented sensitivity and angular resolution. All these missions have laboratory needs for the analysis and interpretation of these datasets.  

There are a number of lessons learned since LAW 2010 and the experience from \textit{Herschel} laboratory efforts that include: 
\begin{itemize}
\item Characterizing the weeds;
\item Searches for new molecules (i.e., the flowers);
\item Consideration of new environments, targets,  and spectroscopic modes;
\item Prioritization of molecules or types of chemistry; and 
\item Laboratory instrumentation and collaboration for the study of new species.
\end{itemize}

\subsubsection{The Era of SOFIA}
Since the end of the \textit{Herschel} mission, the SOFIA airborne observatory has been the only operating FIR mission, with its instruments covering IR wavelengths up to 240 $\mu$m. Studies in the FIR to submm are currently ongoing. New instrumentation, HIRMES (25 - 122 $\mu$m), will open up a new region of study at higher spectral resolution and sensitivity. Open issues include:
\begin{itemize}
\item Continuity of current instruments: Maintenance of current instrument suite, prioritization in upgrades, and relevant laboratory astrophysics.
\item Development of future instruments, 
instrument upgrades with technology advancements or capabilities (e.g. broader wavelength coverage), and corresponding laboratory astrophysics.
\item Dedicated laboratory efforts: New/current instruments have capabilities and wavelength access not routinely covered by established laboratories.  Consideration for dedicated support to expand laboratory efforts for the analysis and interpretation of new data.
\item How best can SOFIA be used to link \textit{Herschel} science products to \textit{JWST} targets?
Consider opportunities that focus on support/complementary observations with future missions and relevant laboratory astrophysics.
\end{itemize}

\subsubsection{Laboratory Needs for the Next Generation Space Telescopes}
Additionally, ESA recently selected the \textit{SPICA} M5 mission concept as one of the three studies for their Cosmic Visions program. One of the four astrophysics mission concept studies for consideration in the 2030s is the \textit{OST}. All these missions have laboratory needs for the analysis and interpretation of these datasets.

\subsubsection{Community Data Needs NOW}
\noindent\includegraphics[width=\textwidth]{Priority_Level_1_ter.png}
\begin{itemize}
\item \textbf{Spectral studies producing rotational line lists, especially focusing on excited vibrational states, isotopologues, and intensities for weaker lines, as well as extending to $>$500 GHz.}  
\item \textbf{Chemical studies to examine the gas/grain interface, bottom-up versus top-down chemistry, fragmentation products, COM formation, chemical budgets for C, N, O, S, etc.}
\item \textbf{FIR studies of ices (optical constants)} - \textit{Herschel}, HIRMES, complement \textit{JWST}
\end{itemize}
\noindent\includegraphics[width=\textwidth]{Priority_Level_2_ter.png}
\begin{itemize}
\item \textbf{Updated cross sections/kinetics/reactions for input to astrochemical models}
\item \textbf{Updated inelastic  collisional data for non-LTE radiative transfer models}
\end{itemize}
\noindent\includegraphics[width=\textwidth]{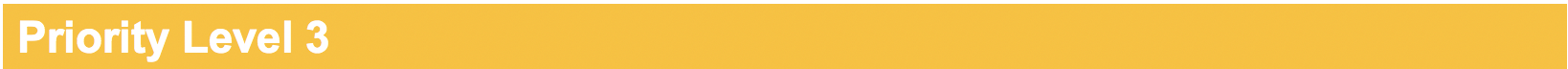}
\begin{itemize}
\item \textbf{Rovibrational studies of larger molecules that can be used for definitive identification of carriers of IR bands that might be observed by \textit{JWST}}
\end{itemize}

\subsubsection{Laboratory Instrumentation/Capability needs}
\noindent\includegraphics[width=\textwidth]{Priority_Level_1_ter.png}
\begin{itemize}
\item \textbf{Higher frequency coverage (e.g., greater than 1 THz to support SOFIA)}
\item \textbf{Higher sensitivity (weak lines, ions, radicals, isotopologues)}
\item \textbf{Faster scanning speeds}
\item \textbf{Automated spectral analysis; analysis tools for complex mixtures}
\end{itemize}
\noindent\includegraphics[width=\textwidth]{Priority_Level_2_ter.png}
\begin{itemize}
\item \textbf{Support for theoretical studies of high-resolution spectroscopy}, particularly to complement experimental work and its interpretation 
\item \textbf{New techniques to study kinetics, gas/grain interfaces, FIR ice studies, large molecule fragmentation}
\end{itemize}


\subsection{X-ray/EUV missions}

The \textit{Chandra X-ray Observatory} and \textit{XMM-Newton} X-ray
gratings have shown the scientific value of high resolution X-ray
spectroscopy around 1 keV, but only for the brightest sources. The
\textit{Extreme Ultraviolet Explorer} and the \textit{Chandra} Low Energy Transmission
Grating have established the spectral connections from ultraviolet to
X-ray. The short-lived \textit{Hitomi} satellite's microcalorimeter spectrum of
the Perseus Cluster demonstrated the potential of high resolution
spectroscopy at higher energies and for extended sources and crowded
fields as well as for point sources, while the CCDs on \textit{Chandra}, \textit{XMM-Newton}, and \textit{Suzaku} observe most of these types of sources at relatively low spectral resolution. The \textit{X-ray Imaging and Spectroscopy
Mission (XRISM)}, and eventually \textit{Athena}, will require improvements to the
spectral models currently being used in the astrophysics community,
including better completeness and higher accuracy.

A survey of the Workshop participants did not reveal clear prioritization; however, experimental photoionization cross section, inner shell ionization experiments, wavelength measurements, and line transition data (electron impact excitation and oscillator strengths) scored on the higher end while photoabsorption and high-temperature DR experiments scored on the lower end. It is interesting to note that almost all of the recommendations were for experimental work, probably reflecting the difficulty of adequately testing or supporting theoretical work. Three priority levels are listed:

\vspace{0.3cm}

\noindent\includegraphics[width=\textwidth]{Priority_Level_1_ter.png}

For X-ray/EUV emitting plasmas, the primary issue is an accurate model of the charge state distributions (CSDs), especially for non-equilibrium environments. Reliable data for the following three processes are key to improving the models:

\begin{itemize}

\item \textbf{Inner-shell ionization of a K- or L-shell electron by either
  photons or electrons} can be an important process in X-ray
  photoionized sources, photoionization of interstellar dust grains,
  and shock-driven gas such as in supernovae. The resulting K- or
  L-shell hole system relaxes via a complicated cascade of
  fluorescence and/or Auger-ejected electrons, affecting both the
  ionization structure and the emitted spectrum.

Inner-shell ionization via photoionization can be studied at the
Photon-ion Spectrometer setup at PETRA III (PIPE) in Hamburg. PETRA III is
presently the brightest third-generation synchrotron light source
worldwide.

\item \textbf{Electron impact ionization (EII)} rates needed to determine the charge state distribution,
especially for higher Z elements, where relativistic effects are
increasingly important. For collisional ionization equilibrium, single
ionization dominates, but multiple ionization is important for
ionizing, non-equilibrium ionization plasmas.

\item \textbf{Low temperature DR} needed to determine
the charge state distribution (CSD) in photoionized sources, including
PNe, X-ray binaries, and AGN. This is difficult because small changes in resonance energies
have dramatic effects on the rate coefficients, so experiments using storage rings are
needed. The Cryogenic Storage Ring (CSR) in Germany might be able to do
all the ground states, but metastable levels introduce uncertainty.

\end{itemize}

\noindent\includegraphics[width=\textwidth]{Priority_Level_2_ter.png}

\begin{itemize}

\item \textbf{High temperature DR} needed to determine the CSD in collisionally
ionized, X-ray emitting plasmas, including stellar coronae, supernova
remnants, and diffuse gas in the ISM. Electron beam
ion trap (EBIT) experiments would be useful to establish satellite
transition wavelengths and relative intensities for X-ray spectra.

\item \textbf{Electron impact excitation rates and oscillator strengths} are needed
for important X-ray and EUV line ratio diagnostics of temperature,
density, non-equilibrium ionization (if occurring), and
abundances. Theory now routinely achieves accuracies to 20\%, but 5-10\%
is needed for many of the diagnostic ratios, especially given typical
signal-to-noise in observations. Complex theoretical calculations,
benchmarked by EBIT spectra, are needed, especially for Fe L-shell ions.

\item \textbf{Theoretical photoionization cross sections at high energies}
  needed for X-ray sources have not been experimentally
  validated. Synchrotron light sources and/or free electron lasers
  such as the Linac Coherent Light Source (LCLS) can provide the
  photoionizing beam, but creating the target and measuring the output
  (such as with an EBIT plasma and associated spectrometer) is a
  significant undertaking. Mobile end-stations at the national
  (or international) facilities would facilitate these experimental
  tests. The Z Facility at Sandia has measured the spectrum of Si
  under photoionization, and the team is currently trying to
  understand discrepancies between the model and experiment.

\item \textbf{Reliable X-ray photoabsorption data for different forms of
  solid- and gas-phase species}, particularly for species containing
  oxygen and iron. Studies of the composition and evolution of the
  cold interstellar medium, e.g. to constrain the mass distribution of
  the progenitor stars. Photoabsorption proceeds either through
  resonant bound-bound transitions or non-resonant bound-free
  transitions, which form distinct spectral features. 

Theoretical calculations of the relevant photoabsorption cross
sections for the O- and Fe-bearing minerals are difficult. Some
progress was made at the Advanced Light Source (ALS) at Lawrence
Berkeley National Laboratory, but this facility has
been decommissioned.

\item \textbf{Charge exchange} calculations do not always agree with each other or
with experiment, particularly as to which \textit{n} and \textit{l} levels
dominate. A concerted collaborative effort between theory and
EBIT experiments is needed to understand what effects are important. Charge
exchange is known to occur in the solar system, as the solar wind
interacts with comets, planetary atmospheres and the
heliopause. Charge exchange might be important in supernova remnants,
the intergalactic medium, and stellar winds; if so, new missions like
\textit{XRISM}, \textit{Athena}, and \textit{Lynx} will need good charge exchange models.

\end{itemize}

\noindent\includegraphics[width=\textwidth]{Priority_Level_3_ter.png}

\begin{itemize}

\item \textbf{Elemental opacities} at conditions appropriate for stellar interiors,
e.g. at the solar interior temperature associated with the tachocline
(i.e., the boundary between the radiative and convective zones). Opacities are
needed for stellar evolution models  and stellar abundances. We note
that the largest source of uncertainty on the age of stars is the
opacity. Opacity measurements at the Z Facility at Sandia National
Laboratory are promising.

\item For all but the strongest emission lines, \textbf{X-ray and EUV wavelengths}
are often too poor (off by $\sim$ 1\%) to allow identifications from
observations. Effort is needed to \textbf{align the transitions used in
astronomy models with experimental data}. Even when experimental data
exist (e.g., NIST data from laser plasmas, EBIT spectra), the rate
calculations used in astrophysics databases do not usually incorporate
these data into the theoretical structures and the theoretical models
that were used to identify the laboratory spectra are not public.

Accurate wavelengths are especially important for the more complex Fe
L-shell ions near 1~keV and for the inner-shell lines to be observed
with microcalorimeters. Absorption spectroscopy requires good models
not only for the
L-shell, but also for the M-shell lines of Fe and other metals, as
these features can be highly blended.

\end{itemize}

\newpage
\subsection{Gamma-ray missions and nuclear/plasma physics}
Participation on this topic was limited at the workshop. However, these are important areas
of laboratory astrophysics which have received little attention from APRA in the past. Participants at the meeting developed a 
list of priorities which is followed here by explicit astrophysical justifications.

\vspace{0.3cm}

\noindent\includegraphics[width=\textwidth]{Priority_Level_1_ter.png}

\begin{itemize}
\item	\textbf{Laboratory measurements and theory for nuclear reaction rates, half-lives, mass measurements, and opacities for neutron-star mergers, kilonova, and multi-messenger astronomy.
} 
\end{itemize}
\noindent\includegraphics[width=\textwidth]{Priority_Level_2_ter.png}
\begin{itemize}
\item	\textbf{Laboratory measurements and theory for nuclear reaction rates and half-lives for key radionuclides.}
\item	\textbf{Laboratory measurements and theory for reaction rates and branching ratios for x-ray bursts.}
\end{itemize}

\vspace{0.5cm}

In addition, the following general recommendations are
given:

\begin{itemize}
\item \textbf{The creation of a center that combines laboratory astrophysics, astrophysics theory, and observations.}
\item \textbf{The development of coordinated software ecosystems in support of NASA missions.}
\end{itemize}

\subsubsection{Neutron-Star mergers, kilonova, multi-messenger astronomy, r-process}
\begin{itemize}
\item	Relevant NASA missions: \textit{JWST}, \textit{Fermi}, \textit{Swift}, \textit{Pan-STARRS}, \textit{HST}, and \textit{Chandra}.
\item	Origin of heavy elements in the universe, $r$-process nucleosynthesis: need nuclear properties including $\beta$-decay half-lives, neutron-capture rates, masses, and fission probabilities. 
\item	Kilonova afterglow due to radioactive decay: need $\beta$-decay properties and atomic opacities.
\item	Sensitivity studies for different astrophysical conditions are important for identifying important properties to be measured.
\item	The properties of $^{132-136}$Cd, $^{134-138}$In, and $^{136-140}$Sn have the largest impact globally. Nuclei near $^{159-161}$Nd have the largest impact locally on the rare earth peak.
\item	Current facilities like the National Superconducting Cyclotron Laboratory and the CARIBU facility at Argonne National Lab provide access to a limited number of $r$-process isotopes. 
\item	In the future the Facility for Rare Isotope Beams (FRIB, 2022) will provide access to the majority of $r$-process nuclei.
\item Constraints on the Equation of State are needed for better understanding of the neutron-star merger events.
\item Measurements or contraints on lanthanide opacities are required for better understanding the kilonova observables.
\end{itemize}

\subsubsection{Radionuclides}
\begin{itemize}
\item	Relevant NASA missions: \textit{INTEGRAL}, \textit{NuSTAR}, Potential future missions such as \textit{LOX}, \textit{HEX-P}, and \textit{e-ASTROGAM}. 
\item	Direct measurement of $\gamma$-rays from long-lived radionuclides, signatures of active nucleosynthesis in the galaxy.
\item	$^{44}$Ti: dependence on reaction rate measurements of the $^{44}$Ti($\alpha$,p)$^{47}$V, $^{45}$V(p,$\gamma$)$^{46}$Cr, $^{17}$F($\alpha$,p)$^{20}$Ne, $^{40}$Ca($\alpha, \gamma$)$^{44}$Ti, $^{40}$Ca($\alpha$,p)$^{43}$Sc, and 3$\alpha$ reactions.
\item	$^{26}$Al: dependence on reaction rate measurements of the $^{26}$Al(n,p)$^{26}$Mg, $^{26}$Al(n,$\alpha$)$^{23}$Na, and $^{22}$Ne($\alpha$,n)$^{25}$Mg reactions.
\item	$^{60}$Fe: dependence on reaction rate measurements of the $^{59}$Fe(n,$\gamma$)$^{60}$Fe, $^{60}$Fe(n,$\gamma$)$^{61}$Fe, $^{22}$Ne($\alpha$,n)$^{25}$Mg, and $^{59}$Fe($\beta^-$)$^{59}$Co.
\item	Some measurements possible with current facilities. Future facilities like FRIB, together with new detector developments, like SECAR, will be able to address these reactions. 
\end{itemize}

\subsubsection{Accreting Neutron stars, X-ray bursts, rp-process}
\begin{itemize}
\item	Relevant NASA missions: \textit{NICER}, \textit{Chandra}, \textit{Swift}, and \textit{Fermi}.
\item	X-ray emission over time for these events varies based on the nuclear physics input.
\item	Accurate description of the $rp$-process depends upon improvements in reaction rate measurements of $^{15}$O($\alpha, \gamma$)$^{19}$Ne - a flow bottleneck, $^{22}$Mg($\alpha$,p)$^{23}$Al - to assess  theoretical rates, (p,$\gamma$) versus (p,$\alpha$) branching at $^{59}$Cu, and the $^{30}$P(p,$\gamma$)$^{31}$S reaction.
\item	Some of the necessary nuclear input is accessible in current facilities with direct or indirect techniques. 
\item	FRIB will provide access to all reactions involved. 
\item	SECAR: the Separator for Capture Reactions is essential for measuring the involved capture reactions directly. Currently under development. 
\end{itemize}

\subsubsection{Plasma Physics and Magnetohydrodynamics Experiments}
\begin{itemize}
\item	Relevant NASA missions: \textit{Spitzer}, \textit{Swift}, \textit{JWST}, and \textit{HST}.
\item	Experiments on molecular dynamics of dust particles in astrophysical plasmas. 
\item	Experiments probing dust-gas-plasma interaction and dust growth and breakup.
\end{itemize}

\subsection{Databases, Codes, and Archives}

Databases of atomic and molecular data, spectroscopic features, and other physical properties and quantities, and the software suites that produce or manipulate them, are of key importance to most areas of laboratory astrophysics. The maintenance and curation of these assets, obtained through APRA funding and other sources, is a central issue to the field.

Thus, LAW 2018 dedicated a half-day of presentations and devoted two separate breakout sessions to the various issues relating to laboratory astrophysics codes (more properly, ``software instruments"), databases, and the long term preservation and archiving of both. Several critical questions emerged, and considerable effort was expended in attempting to identify possible solutions.

\begin{enumerate}

\item How do we maintain and preserve valuable and well-established laboratory astrophysics codes and databases once the original funding that supported them withers away, or once their lead authors and other key individuals retire or move onto other projects?

\item What standards or best practices exist, or should exist, concerning the reliability of publicly released codes, especially those initiated or funded from NASA programs?

\item What should the role of the three NASA-funded astrophysics data archives (HEASARC, IRSA, MAST) and other associated NASA-funded centers be (e.g., in code preservation)?

\item What other related issues should NASA and the community be bearing in mind, as we enter the next decade of laboratory astrophysics?
\end{enumerate}

Here, we address each of these areas in turn.

\subsubsection{Maintenance and Preservation of Key Codes and Databases}

\begin{itemize}
\item NASA Astrophysics supports the maintenance of some existing atomic, molecular, nuclear, and other databases through ADAP, mission budgets, and the like. NIST supports the maintenance of other relevant databases for the science community, some specifically tuned to the needs of astrophysics, others less so. Within the U.S., some well-known examples include the JPL Molecular Spectroscopy database, the Ames PAH IR Spectroscopic Database (PAHdb), the AtomDB (Atomic Database for Astrophysicists) project, NRAO's Splatalogue, the Chianti atomic database, the HITRAN molecular spectroscopic database, and many more. We also note the popularity and importance of various complementary European resources, including (but obviously not limited to) CDMS (the Cologne Database for Molecular Spectroscopy), VAMDC (Virtual Atomic and Molecular Data Center) Consortium portal, and KIDA (Kinetic Database for Astrochemistry). Many spectral codes are of central significance to astrophysics, including Cloudy, XSTAR, pyAtomDB, and SPEX, and a number of fitting packages provide valuable interpretive insight to laboratory astrophysics data, including XSPEC (maintained at GSFC), Sherpa (maintained by CXC), and ISIS (maintained at MIT). We could name equally long lists of national and international capabilities in laboratory nuclear astrophysics, stellar evolution, and hydrodynamics. Some of these resources have adequate funding levels, but many others are surviving year-to-year, receiving intermittent funding from a variety of sources, or are in danger of interruption due to loss of funding or critical personnel. In such an environment, we consider it critical that NASA establish a mechanism to ensure the uninterrupted continuation of support for the databases and tools with key relevance to current and future missions, and to NASA's science goals as defined by the Decadal Surveys and subsequently laid out in NASA Science Plans.

\item Software sustainability is an important issue. Sustainability is a common topic at software meetings large and small, within and outside astrophysics, and there are various models for beginning software projects to draw upon. Software that is community-driven from its inception has a much better chance of continuing to be community-supported in the long-term, and the presence or absence of such community support could be considered a coarse metric of the importance of that particular software package: popular packages will tend to receive a sustainable level of community support. 

\item However, the laboratory astrophysics field can be considered small compared to other sectors of the astrophysical software community, and many critical software packages have been developed by a single expert or small team of experts (the so-called ``ace programmer" model). This then leads to the well-recognized succession issue. It is frequently difficult to find a replacement for such central figures once they retire or move onto other duties, and often even more difficult to identify a funding source for such a replacement. Ace programmers tend to perform their software development as part of (or instead of) a research career. Ideally, one might hope that the community could jump in and crowdsource the development into the future, but in reality this approach is much more difficult with a large and complex integrated code (such as CLOUDY, XSTAR, XSPEC) than with a set of libraries (such as exist in NumPy, AstroPy). Thus, the transition of specific packages from the ``ace programmer" model to the ``community-supported" model may realistically be challenging for most of the workhorse codes discussed at LAW 2018. Note that even where a code or software suite might be used and/or cited by thousands of researchers annually, it may still not have sufficient community among programmers to enable the switch to a crowd-sourced support model.

\item A natural place to host codes is an online repository; popular current repositories include GitHub and SourceForge. While this is a perfectly valid approach, and one would hope that such repositories will be long-lived, there is no guarantee that such codes will be maintained and updated to work on new versions of various operating systems, and no guarantee that the codes will be documented or that help can be obtained to install and run them, or correctly interpret the results. In many cases, writing the actual code is only a quarter of the job; thorough testing, effective documentation, and other necessary activities take up the bulk of the effort.

\item For the above reasons, and in the longer term, the existing astrophysics archives and facilities are obvious places to curate and maintain vital laboratory astrophysics databases and codes. In some cases the major archives might be the appropriate resting places: the HEASARC for laboratory astrophysics databases relevant to the high energy astrophysics communities (X-ray, gamma-ray); MAST for optical and UV databases, and IRSA/IPAC for IR-submm databases. The staff of these archives have extensive expertise with astrophysics in their wavelength area, and many laboratory astrophysics databases and codes would be a natural fit to these archives. Other databases and resources might well be better retained by their originating institutions, but the funding of these resources would need to be put on a more stable basis.

\begin{itemize}
\item These additional responsibilities would obviously come with a cost. That cost would depend upon the level of maintenance and portability expected, the continuing expertise required on staff with the relevant archive or location, and other criteria.

\item Metrics that might be used to determine whether sufficient community exists to fold existing databases and codes into one of the archive centers might include the number of registered users, the degree of acknowledgment of the databases/codes in refereed journal papers, and the number of citations to papers that specifically describe the codes and databases under discussion.

\end{itemize}
\end{itemize}

\noindent\includegraphics[width=\textwidth]{Priority_Level_1_ter.png}

\begin{itemize}
\item Having carefully considered all of the above, \textbf{we encourage NASA to develop new funding resources to support critical, sustainable, community-driven software and infrastructures at a variety of scales, in support of current and future NASA missions.} Within this vision, the new program should have the overarching goal of transforming innovations in research into sustained software and data resources that become an integral part of NASA's cyberinfrastructure. This program would be cross-foundational, generating and nurturing the interdisciplinary processes required to support the entire software and data life cycle. Such a program would also encourage vibrant partnerships among academic institutions, government laboratories, and industry.

\end{itemize}

\subsubsection{Standards and Best Practices for Publicly-Released Codes/Data}

\begin{itemize}
\item Strict version control of code, related calibration files, documentation, and other ancillary materials are essential.

\item Many existing software instruments, both NASA-funded and community-driven, include test suites that run on a regular basis on all supported platforms and operating systems. The results of these tests are either monitored internally by the software developers, or posted to public web sites. In either case, such test suites enable straightforward validation and verification of code performance, and we recommend they should be built into all new packages from the start.
\end{itemize}

\subsubsection{Role of Existing Centers}

\begin{itemize}
\item As noted above, for long-term curation of laboratory astrophysics codes and databases, the existing astrophysics archives and facilities may be obvious places to curate and maintain vital laboratory astrophysics databases and codes. The expertise of these archival research centers in the wavelength regime of interest make them good candidate locations for the maintenance of laboratory astrophysics software assets in many cases.
\end{itemize}

\subsubsection{Other Issues Related to Codes, Databases, and Archives}

\begin{itemize}
\item There is growing expertise in the astronomical community about issues related to big data, machine learning, and neural networks. Such expertise may be of interest to the laboratory astrophysics community.

\item Many academic journals in astrophysics, quantum chemistry, and other fields of interest now ask authors to deposit data, codes, and documentation along with the texts of their papers. Such deposits form an additional route of providing long term access to such materials. However, codes archived at journals will probably not be maintained and updated in the same way as codes located at an astrophysics archive or community repository. We view the role of journals in archiving software and databases to be valuable, but insufficient.
\end{itemize}

\section{Other General Issues for Consideration}

\subsection{Facilities}

Facilities needed for X-ray and EUV experiments include EBITs, cryogenic storage rings, synchrotron light sources (e.g., PETRA III), the free electron laser facility LCLS, and the Z Machine at Sandia National Laboratory, as well as facilities relevant to other wave bands. Some of these facilities have unique capabilities. We recommend that NASA do an assessment of facility needs, identify DoE or international facilities that could be used for laboratory astrophysics experiments, and meet with the other agencies to discuss mutual needs.
The community also strongly supports NASA implementing a program that would track instruments that are underutilized and facilitate their use. For European facilities, should there be a joint laboratory astrophysics program for \textit{Athena}? A risk assessment might be in order if certain capabilities do not currently exist. At the larger facilities such as the synchrotrons, mobile end-stations could be extremely useful to capitalize on opportunities for laboratory astrophysics.

\subsection{Other Topics}

\begin{itemize}
\item NASA could initiate and support code comparisons, e.g. photoionization codes or collisional ionization codes, so that accuracy and completeness can be assessed by the community. These have generally been organized by individuals (Lorentz Center) or other institutions like the Non-LTE Code Comparison workshop (NIST).

\item NASA could develop a program for visiting students, so that laboratory astrophysics could work collaboratively with astrophysicists and vice versa.

\item NASA should determine whether junior laboratory astrophysicists are applying for the prestigious NASA fellowships (Hubble, Einstein, Sagan). If not, recruitment efforts would be worthwhile and the application evaluation committees should be advised that such applications fit within the goals of the fellowship programs. 

\item Should the community, as proposed above, engage in the development of laboratory astrophysics centers? The danger is that this would reduce the limited available funds  for individual laboratory astrophysics investigators; a topic discussed at LAW 2010.

\item The idea that missions should support through their budgets their own laboratory astrophysics data needs was discussed at LAW 2010 and LAW 2018 and endorsed here. Missions do not necessarily have the long-term vision needed to support decades-long efforts to improve the laboratory astrophysics data. On the other hand peer review may not necessarily produce the most relevant data.

\item NASA should host annual workshops on specific topics to promote communication between laboratory astrophysicists and observers and such workshops might be tied to specific missions. Such workshops might be organized in collaboration with appropriate divisions in the American Astronomical Society (AAS) and the American Chemical Society (ACS).
\end{itemize}

\section{Outcomes from LAW 2010 Recommendations}

Finally, we briefly review the successes that derived from recommendations made in the
LAW 2010 White Paper as well as recommendations that were not acted upon.

\begin{itemize}
\item {\em Restoring APRA funding to the baseline funding level of FY2006.}  Funding levels for APRA laboratory astrophysics science was \$3.5M, \$3.5M, and \$4M for FY2008, FY2010, and FY2017. As a consequence this goal was only partially realized, but it is noted that steady increases have occurred since FY2014.  
\item {\em Implementing the Astro2010 Decadal Survey recommendations to grow the APRA program support and to provide mission support for laboratory astrophysics.} It was recommended to add \$2M/year beyond the baseline, which was partially successful considering FY2010 as the baseline. Laboratory astrophysics support from missions was partially successful as a number of pure laboratory astrophysics proposals were awarded by \textit{HST} and \textit{Spitzer} since FY2010. Upcoming \textit{JWST} Guest Observer proposals will now allow up to 15\% of effort for laboratory astrophysics. We encourage other current and future mission observation cycle calls to include a laboratory astrophysics component.
\item {\em Establishing a series of new initiatives in order to revitalize, grow, and ensure the future of laboratory astrophysics.} In 2011, the APRA solicitation included a call for team/group grand-challenge projects on carbon in the universe and a proposal was funded. However, this was not repeated in subsequent calls. Other initiatives such as \textit{laboratory astrophysics specific} programs for graduate student and postdoctoral fellowships, young-investigator type proposal programs, and technology development and instrumentation programs were not acted upon. Though young researchers in laboratory astrophysics may be supported through the established NASA Graduate Research Fellowship and Postdoctoral Programs.
\item {\em Developing an appropriate mechanism to ensure the long-term viability of laboratory astrophysics databases.} While ADAP has a component to support laboratory astrophysics databases, these provide short-term, 3-year funding. Long-term viability still remains a concern as highlighted in this report.
\item {\em Continuing to sponsor this quadrennial series of Laboratory Astrophysics Workshops, the next meeting of which should take place in the 2014-2015 period.} This was eventually realized by the convening of LAW 2018. Further, a yearly APRA Laboratory Astrophysics PI meeting, of more limited scope, was established in 2015 or thereabouts, while a more focused meeting, the Workshop on Molecular Spectroscopy in the Era of Far-Infrared Astronomy, was held in 2012. A NASA LAW meeting in 2022 is recommended.
\end{itemize}

Not surprisingly, the vast majority of the laboratory astrophysics data needs itemized in this document closely match those listed in the
LAW 2010 White Paper. While progress has been made on many data-need fronts, the lack of a consensus priority ranking may have led to only modest incremental advances. 

\section{Conclusions}

LAW 2018 brought together 95 scientists including astronomers, 
modelers, chemists, physicists, and mission scientists with the goal of reviewing the status of the laboratory astrophysics field and looking 
into the current and future needs 
to support NASA astrophysics missions. Unlike previous LAW meetings,
our charter was to evaluate and \textit{prioritize} laboratory astrophysics data required to meet the scientific goals of current and approved future NASA missions. 
The detailed prioritizations given in the body of the Report are the result of a community effort and were influenced by near-term (i.e., \textit{JWST}, \textit{XRISM}) and currently operating (e.g., SOFIA, \textit{HST}, \textit{Chandra}, ...) missions. 
However, laboratory astrophysics needs will certainly evolve as unforeseen issues occur or dramatic new discoveries are made.
The prioritizations listed in this report can serve as guidelines for proposal writers and will be used by NASA program officers to help guide NASA investments. However, the Report is not meant to override appropriate peer review of research proposals. This LAW Report will help justify and advocate for laboratory astrophysics within NASA and will help prepare for the decadal survey.

We end by reemphasizing some over-arching recommendations that can affect both the creation
and long-term maintenance of laboratory astrophysics data.

\begin{itemize}

\item The concept of an Institute for Laboratory Astrophysics, either as a stand-alone entity, part of a broader astronomy center, or a network of smaller facilities, has been discussed for the past two decades. Limited funds in any economic environment would place support of individual investigators in jeopardy. Known attempts through single agency programs have been unsuccessful. The community may wish to think about a multi-agency approach. 

\item We encourage NASA to develop new funding mechanisms to support sustainable laboratory astrophysics databases as critical, community-driven software instruments with supporting infrastructures, at a variety of scales.
\item We encourage NASA to develop a complete list of, and then coordinate with, experimental user facilities at other US agencies and international institutions to ensure continued access for laboratory astrophysics relevant experiments. Loss of access to key facilities could be detrimental to NASA scientific goals.
\item We encourage the laboratory astrophysics community, in coordination with NASA and other stakeholders, to prepare white papers for the 2020 Decadal Survey for Astronomy and Astrophysics.
 
\end{itemize}

\newpage
\section{Appendix: Decadal survey preparation}

The upcoming Decadal Survey (DS) for Astronomy and Astrophysics was taken up by the Workshop participants during a break-out session at LAW 2018, with additional opportunity for community input following the meeting. The DS is sponsored by the National Academy of Sciences (NAS), with support from the Committee for Astronomy and Astrophysics (CAA). The DS report is submitted to the U.S. federal agencies, DoE, NASA, and NSF, as community-based input for funding and investment in astronomy. Previous DS reports and the most recent mid-decadal review are available from the NAS web site. 

Since 1972, each DS report has included statements pertaining to the field of laboratory astrophysics, highlighting critical data needs for the future. It is anticipated that the upcoming DS report will also include highlights of critical laboratory data needs. The term laboratory astrophysics will in this discussion include data that originates through laboratory experiment, theory, and calculation. Although NASA sponsored the Workshop, and financially supports some ground-based astronomy projects, the discussions of the DS encompassed space- and ground-based astronomy observations, as well as theoretical investigations.

The Workshop discussion reviewed the expected milestones and time frame for the DS report, which culminates with the final report in calendar year 2020. This information was taken from the presentation of CAA Co-Chairs Drs. M. Rieke and S. Ritz, as presented at the session on DS Planning at the January 2018 AAS meeting. The laboratory astrophysics community will have additional opportunities to present input to the DS, either through white papers organized by the AAS or through individual contributions, and the discussions at LAW were considered as preparation for future discussions and actions. It was recognized by the Workshop participants that there should be continual communication originating from the white paper coordinating committee(s) to the community for their input. These organizations can include the AAS, the ACS, and the American Physical Society (APS). 

The Workshop took up issues of laboratory astrophysics needs for high priority science for the future, as well as the state of the laboratory astrophysics community. The science discussion sessions were divided into the wavebands gamma-ray, x-ray, UV-optical-NIR, MIR, and FIR/sub-mm, with each waveband prioritizing its laboratory astrophysics data needs. Since the timing of the Workshop is close to the start of the DS process, and given the length of time to realize research results funded through typical agency funding opportunities, the waveband prioritizations can be considered to apply during the period covered by the upcoming DS. These waveband prioritizations are presented in the main body of this LAW report. 

The field of laboratory astrophysics has expanded greatly from beyond the determination of wavelengths and oscillator strengths of atomic and molecular transitions that dominated early DS science. Improvements in x-ray instrument technology, with micro-calorimeter detectors receiving notable attention at this Workshop, offer stark improvements in resolution and sensitivity, which will require a concerted effort to interpret new high quality spectra. Likewise, the interpretation of exoplanet atmosphere spectra will require further work in molecular spectroscopy. 
While these areas remain important to interpreting new astronomy data, there has been a continuing expansion of research in areas characterized as astrochemistry, astrobiology, and nuclear astrophysics, the latter highlighted by the recent kilonova observation. 

The discussion on the “health of the field” included several areas that are considered critical to the laboratory astrophysics community:
 
\begin{itemize}
\item Career paths for young investigators need to be better defined, with an effort made to retain talented researchers.
\item Aging of the field, as many university professors may be close to retirement and there was concern that these positions may not be refilled.
\item Diversity within the field, with particular emphasis on the underrepresentation of women and minorities.
\item The Laboratory Astrophysics community needs to be better portrayed to attract young STEM students and provide insight for the public, even though the subject of astronomy retains the public's fascination. 
\end{itemize}

In general, funding remains a concern to the laboratory astrophysics community both in terms of sustainability of facilities and researchers, and attracting and retaining new talent. Decreased funding within DoE raises concerns that other agencies (NASA, NSF) will feel increased funding pressure.  This concern might be extended to other agencies, for example NIST support of laboratory astrophysics related activities (databases). Suggestions were made for other funding mechanisms outside the typical small research grant opportunities found within NASA and NSF. These included directed funding within space mission budgets, and increased support for workshops, Centers of Excellence, and Grand Challenges. NASA APRA did fund a Grand Challenge project (Carbon in the Universe, 2012), which included international partnerships and resulted in numerous published studies.

International partnerships are critical to conducting science. With relatively small numbers of people in each subfield of laboratory astrophysics, the international community needs to work together to avoid the unnecessary duplication of work and use the strengths of each institution to maximize output. Addressing certain research problems will rely on leveraging multiple facilities and institutions in the U.S. and abroad. Many NASA missions carry ESA instruments, and NASA participates in foreign agency missions such as the Japanese Aerospace Exploration Agency. Funding opportunities for laboratory astrophysics can also be identified, and realized at relatively little cost. 

While LAW 2018 took up the need for new facilities, there was also discussion that there be a means for obtaining and paying for time on foreign or private facilities, outside the regular facilities that are funded by NASA, NSF or the DoE. It is recommended that these agencies take up the discussion among themselves and with international partners to allow U.S. or specific agency funded participation in the use of foreign laboratory astrophysics facilities. Clearly, the laboratory astrophysics community should begin developing these and related topics to craft DS white papers as soon as possible.

\newpage
\section{Acronym List}

\textbf{AAS}: 		American Astronomical Society\\
\textbf{ACS}: 		American Chemical Society\\
\textbf{ADAP}:       Astrophysics Data Analysis Program\\
\textbf{AGN}:		Active Galactic Nuclei\\
\textbf{ALS}:		Advanced Light Source\\
\textbf{APRA}:      Astrophysics Research and Analysis Program\\
\textbf{APS}: 		American Physical Society\\
\textbf{Athena}: 	Advanced Telescope for High Energy Astrophysics\\
\textbf{AtomDB}:	Atomic Database for Astrophysicists\\
\textbf{Chandra}: 	Chandra X-ray Observatory\\
\textbf{CDMS}:		Cologne Database for Molecular Spectroscopy\\
\textbf{COMs}:		Complex Organic Molecules\\
\textbf{CSD}:		Charged State Distribution\\
\textbf{CSR}:		Cryogenic Storage Ring\\
\textbf{DoE}:		Department of Energy\\
\textbf{DR}:		Dielectronic Recombination\\
\textbf{DS}:		Decadal Survey\\
\textbf{e-ASTROGAM}: enhanced-ASTROGAM\\
\textbf{EBIT}:		Electron Beam Ion Trap\\
\textbf{EII}:		Electron Impact Ionization\\
\textbf{Fermi}:		Fermi Gamma-ray Space Telescope\\
\textbf{FIR}: 		Far Infrared\\
\textbf{FRIB}: 		Facility for Rare Isotope Beams\\
\textbf{FUSE}: 		Far UV Spectroscopic Explorer\\
\textbf{GALEX}:		Galaxy Evolution Explorer\\
\textbf{HabEx}:		Habitable Exoplanet Imaging Mission\\
\textbf{HEASARC}:	High Energy Astrophysics Science Archive Research Center\\
\textbf{HITRAN}: High-resolution Transmission molecular Absorption database\\
\textbf{Herschel}:	Herschel Space Observatory\\
\textbf{HEX-P}:		High Energy X-ray Probe\\
\textbf{HST}:		Hubble Space Telescope\\
\textbf{INTEGRAL}:	INTErnational Gamma-Ray Astrophysics Laboratory\\
\textbf{IPAC}: 		Infrared Processing and Analysis Center\\
\textbf{IR}:		Infrared\\
\textbf{IRAS}:		Infrared Astronomical Satellite\\
\textbf{IRSA}:		(NASA/IPAC) Infrared Science Archive\\
\textbf{ISO}:		Infrared Space Observatory\\
\textbf{IUE}:		International Ultraviolet Explorer\\ 
\textbf{JWST}:		James Webb Space Telescope\\
\textbf{KIDA}:		Kinetic Database for Astrochemistry\\
\textbf{LAW}:		Laboratory Astrophysics Workshop\\
\textbf{LCLS}:		Linac Coherent Light Source\\
\textbf{LOX}:		Lunar Occultation Explorer\\
\textbf{LTE}: Local Thermodynamic Equilibrium\\
\textbf{LUVOIR}:	Large UV Optical Infrared Surveyor\\
\textbf{MAST}:		Mikulski Archive for Space Telescopes\\
\textbf{MIR}:		Mid-infrared\\
\textbf{NICER}:		Neutron Star Interior Composition Explorer\\
\textbf{NIST}:      National Institute of Standards and Technology\\
\textbf{NSM}:		Neutron Star Merger\\
\textbf{NuSTAR}:	Nuclear Spectroscopic Telescope Array\\
\textbf{OST}:		Origins Space Telescope\\
\textbf{PAH}:		Polycyclic Aromatic Hydrocarbon\\
\textbf{PAHdb}:		NASA Ames PAH IR Spectroscopic Database\\
\textbf{PANH}:		Nitrogenated Polycyclic Aromatic Hydrocarbon \\
\textbf{Pan-STARRS}: Panoramic Survey Telescope and Rapid Response System\\
\textbf{PNe}:		Planetary Nebulae\\
\textbf{SECAR}:		Separator for Capture Reactions \\
\textbf{SPICA}:		Space Infrared Telescope for Cosmology and Astrophysics\\
\textbf{Spitzer}:	Spitzer Space Telescope\\
\textbf{SOFIA}:		Stratospheric Observatory For Infrared Astronomy\\
\textbf{Swift}:		Neil Gehrels Swift Observatory\\
\textbf{TESS}:		Transiting Exoplanet Survey Satellite\\
\textbf{UV}:		Ultraviolet\\
\textbf{WISE}:		Wide-field Infrared Survey Explorer\\
\textbf{XMM-Newton}: X-ray Multimirror Mission - Newton\\
\textbf{XRISM}:		X-ray Imaging and Spectroscopy Mission\\

\end{document}